\documentclass[12pt,preprint]{aastex}
\usepackage[]{natbib}
\usepackage{graphicx, amsmath}
\begin{document}

\title{H$\alpha$ Star Formation Rates of z $>$ 1 Galaxy Clusters in the IRAC Shallow Cluster Survey}
\author{Gregory R. Zeimann \altaffilmark{1,2}, S. A. Stanford \altaffilmark{1,3}, Mark Brodwin \altaffilmark{4}, Anthony H. Gonzalez \altaffilmark{5}, Conor Mancone\altaffilmark{5}, Gregory F. Snyder\altaffilmark{6}, Daniel Stern \altaffilmark{7}, Peter Eisenhardt \altaffilmark{7}, Arjun Dey \altaffilmark{8}, and John Moustakas \altaffilmark{9}}
\altaffiltext{1}{Department of Physics, University of California, One Shields Avenue, Davis, CA 95616}
\altaffiltext{2}{Department of Astronomy \& Astrophysics, The Pennsylvania State University, University Park, PA 16802}
\altaffiltext{3}{Institute of Geophysics and Planetary Physics, Lawrence Livermore National Laboratory, Livermore, CA 94550}
\altaffiltext{4}{University of Missouri-Kansas City, Kansas City, MO, 64110}
\altaffiltext{5}{Department of Astronomy, University of Florida, Gainesville, FL 32611}
\altaffiltext{6}{Harvard-Smithsonian Center for Astrophysics, 60 Garden Street, Cambridge, MA 02138}
\altaffiltext{7}{Jet Propulsion Laboratory, California Institute of Technology, Pasadena, CA 91109}
\altaffiltext{8}{NOAO, 950 North Cherry Avenue, Tucson, AZ 85719}
\altaffiltext{9}{Department of Physics and Astronomy, Siena College, 515 Loudon Road, Loudonville, NY 12211}

\begin{abstract}
We present {\it Hubble Space Telescope} near-IR spectroscopy for 18 galaxy clusters at 1.0 $<$ $z$ $<$ 1.5 in the IRAC Shallow Cluster Survey.  We use Wide Field Camera 3 grism data to spectroscopically identify H$\alpha$ emitters in both the cores of galaxy clusters as well as in field galaxies.  We find a large cluster-to-cluster scatter in the star formation rates within a projected radius of 500 kpc, and many of our clusters ($\sim$60\%) have significant levels of star formation within a projected radius of 200 kpc.  A stacking analysis reveals that dust reddening in these star-forming galaxies is positively correlated with stellar mass and may be higher in the field than the cluster at a fixed stellar mass.  This may indicate a lower amount of gas in star-forming cluster galaxies than in the field population.  Also, H$\alpha$ equivalent widths of star-forming galaxies in the cluster environment are still suppressed below the level of the field.  This suppression is most significant for lower mass galaxies ($\log$ M$_{*}$ $<$ 10.0 M$_{\odot}$).  We therefore conclude that environmental effects are still important at 1.0 $<$ $z$ $<$ 1.5 for star-forming galaxies in galaxy clusters with $\log $ M$_{*}$ $\lesssim$ 10.0 M$_{\odot}$.  
\end{abstract}
\label{sec:abs}

\section{Introduction}
\label{sec:intro}

Growing out of the cosmic web, galaxy clusters provide insights into the formation and growth of large-scale structure as well as the physics that drives galaxy evolution.  Even at $z$ $\gtrsim$ 1 galaxy clusters harbor a high density of old, massive stellar populations (\citealp{eisenhardt2008,snyder2012}), providing an early glimpse of the stellar mass build-up in rich, highly biased environments (e.g., \citealp{mancone2010,mancone2012,lemaux2012,snyder2012,rudnick2012}). 

Locally, there is a well established relation between environment or density and star formation (e.g. \citealp{gomez2003}).  The centers of low-redshift clusters show no evidence of significant ongoing star formation and consist of mostly massive, red galaxies.  The commonly used model for the formation of these massive cluster galaxies is a short, intense burst of star formation at high redshift ($z$ $\sim$ 3), followed by passive evolution (e.g. \citealp{stanford1998,eisenhardt2008}).  However, these simple models are ruled out by observations of clusters at $1 < z < 2$, which suggest that continuous and ongoing star formation is occurring at these redshifts (\citealp{snyder2012}).  The cessation or suppression of that star formation in cluster cores may be caused by a variety of environmental effects (e.g. strangulation, ram-pressure stripping, and galaxy harassment; \citealp{larson1980,moore1999}), and the epoch at which these effects become important is still unknown.

Studies of the star formation rate - local density relation at high redshift ($z$ $>$ 1) have yielded varying results.  There is evidence that in some clusters the environmental effect on star formation is not yet significant (\citealp{hilton2010,tran2010,brodwin2013,alberts2013}).  Other clusters seem to already have environmental effects in place at $z$ $\lesssim$ 1.4, with star formation ceased in the core (\citealp{tanaka2009,grutzbauch2012,muzzin2012}).  This may reflect a diversity of intracluster media and dynamical histories for clusters currently studied at $z$ $>$ 1, which is plausible given that they are selected from a variety of methods and cover a range of cluster masses (M$_{200}$ $\sim$ 0.8 - 9 $\times$ 10$^{14}$ M$_{\odot}$).    

A large statistical sample of uniformly-selected galaxy clusters at $z$ $>$ 1 can provide an ideal testbed for star formation in cluster cores and examining the role of environment in regulating star formation.  The stellar mass-selected IRAC Shallow Cluster Survey (ISCS; \citealp{eisenhardt2008}) includes more than 20 spectroscopically confirmed clusters at $z$ $>$ 1 (\citealp{stanford2005, brodwin2006, elston2006, eisenhardt2008, brodwin2011, brodwin2013}).  We observed 18 of these clusters with the {\it Hubble Space Telescope's} Wide Field Camera 3 (HST/WFC3) grism, which allows the spectral identification of H$\alpha$ emission for all objects in the dense cores of these clusters.

We present the ISCS in \S2, including all relevant data to this work, the data reduction and H$\alpha$ measurements in \S3 and \S4, respectively.  We present the physical implications of the H$\alpha$ star formation rates in \S5. We use a \citet{chabrier2003} initial mass function and a WMAP7+BAO+$H_0$ $\Lambda$CDM cosmology (\citealp{komatsu2011}): ${\Omega}_{M}$ = 0.272, ${\Omega}_{\Lambda}$ = 0.728, and $H_0$ = 70.4 km s$^{-1}$ Mpc$^{-1}$.

\section{IRAC Shallow Cluster Survey}
\label{sec:obs}

\subsection{Survey}

The ISCS identified cluster candidates over an area of 7.25 deg$^2$ in the Bo\"{o}tes field of the NOAO Deep Wide-Field Survey (NDWFS; \citealp{jannuzi1999}). The clusters were identified as 3-D spatial overdensities (RA, Dec, and photometric redshift) using accurate optical/IR photometric redshifts (\citealp{brodwin2006}) calculated for the 4.5 $\mu$m flux-limited (8.8 $\mu$Jy at 5$\sigma$) catalog of the IRAC Shallow Survey (ISS; \citealp{eisenhardt2004}). The ISCS compiled a catalog of over 300 cluster candidates spanning 0.1 $<$ $z$ $<$ 2, including more than 100 at $z$ $>$ 1.  More than 20 clusters at 1 $<$ $z$ $<$ 1.5 have been spectroscopically confirmed to date (\citealp{stanford2005, brodwin2006, elston2006, eisenhardt2008, brodwin2011, brodwin2013}).  A variety of mass proxies including X-ray luminosity and temperature, weak-lensing, and near-IR luminosity have been measured for a subset of the ISCS clusters (\citealp{eisenhardt2008,brodwin2011,jee2011}). These indicate masses in the range of M$_{200}$ $\sim$ $(1-5) \times 10^{14} M_{\odot}$, consistent with the mean mass obtained by comparing the clustering of the ISCS cluster sample with N-body simulations (\citealp{brodwin2007}).

\subsection{Optical/Near-IR/IRAC Imaging}

Optical data from the NDWFS (B$_{W}$RI; \citealp{jannuzi1999}) are available for all ISCS clusters.  Aperture-corrected 4$\arcsec$ fluxes were used to match the larger PSFs of the {\it Spitzer}/IRAC photometry (see \citealp{brodwin2006} for more details).  Recently, we obtained near-IR (NIR) data from the NOAO Extremely Wide-Field Infrared Imager (NEWFIRM) in J, H, and K$_{s}$ that cover all of the NDWFS. 

The {\it Spitzer} Deep, Wide-Field Survey (SDWFS; \citealp{ashby2009}) increased the ISS depth by a factor of four in exposure time.  Combined with PSF-matched NDWFS optical catalogs, these data were used to compute new photometric redshifts for the full 4.5 $\mu$m flux-limited SDWFS sample (5.3 $\mu$Jy at 5$\sigma$). 

\subsection{{\it HST} Spectroscopy/Imaging}

Both high-resolution near-IR (NIR) imaging and NIR slitless spectroscopy (GO proposal ID 11597) were obtained for 18 $z$ $>$ 1 clusters in the ISCS sample using the {\it HST} Wide Field Camera 3 (WFC3; \citealp{kimble2008}).  The program targeted the 18 high-redshift clusters with the G141 grism which has a throughput greater than 10$\%$ in the range of 1.08 - 1.69$\mu$m and a resolution of  93 \AA, sufficient to securely identify cluster members with a typical redshift accuracy of ${\sigma}_z$ $\approx$ 0.01.  The total integration for each target with the grism was 2011s and was comprised of four individual, dithered exposures.  Accompanying each dithered grism exposure was a 103s direct image with the F160W filter which was used for source identification and wavelength calibration of the spectra.  The field of view for both the grism and the direct image is 136$\arcsec \times 123\arcsec$ ($\sim$ 1.1 Mpc $\times$ 1.0 Mpc at z $=$ 1).  Five of the 18 targets have multiple visits due to the fact that the initial pointings missed the cluster centers by 30-80$\arcsec$.  Redshifts were obtained for all pointings, but for uniformity, the analysis that follows only uses the pointing closest to the cluster center.      

A variety of other programs (GO proposal IDs 10496, 11002, 11663) provided optical data for a subset of the clusters with the Advanced Camera for Surveys (\citealp{ford1998}) and Wide Field Planetary Camera 2 (\citealp{holtzman1995}) in filters F775W, F850LP, and F814W.  Pseudo-color images (F775W+F850LP+F160W) for four clusters are shown in Figure 1.

\begin{figure*}[htp] 
\setlength\fboxsep{0.0pt}
\setlength\fboxrule{0.0pt}
\fbox{\includegraphics[width=.45\textwidth]{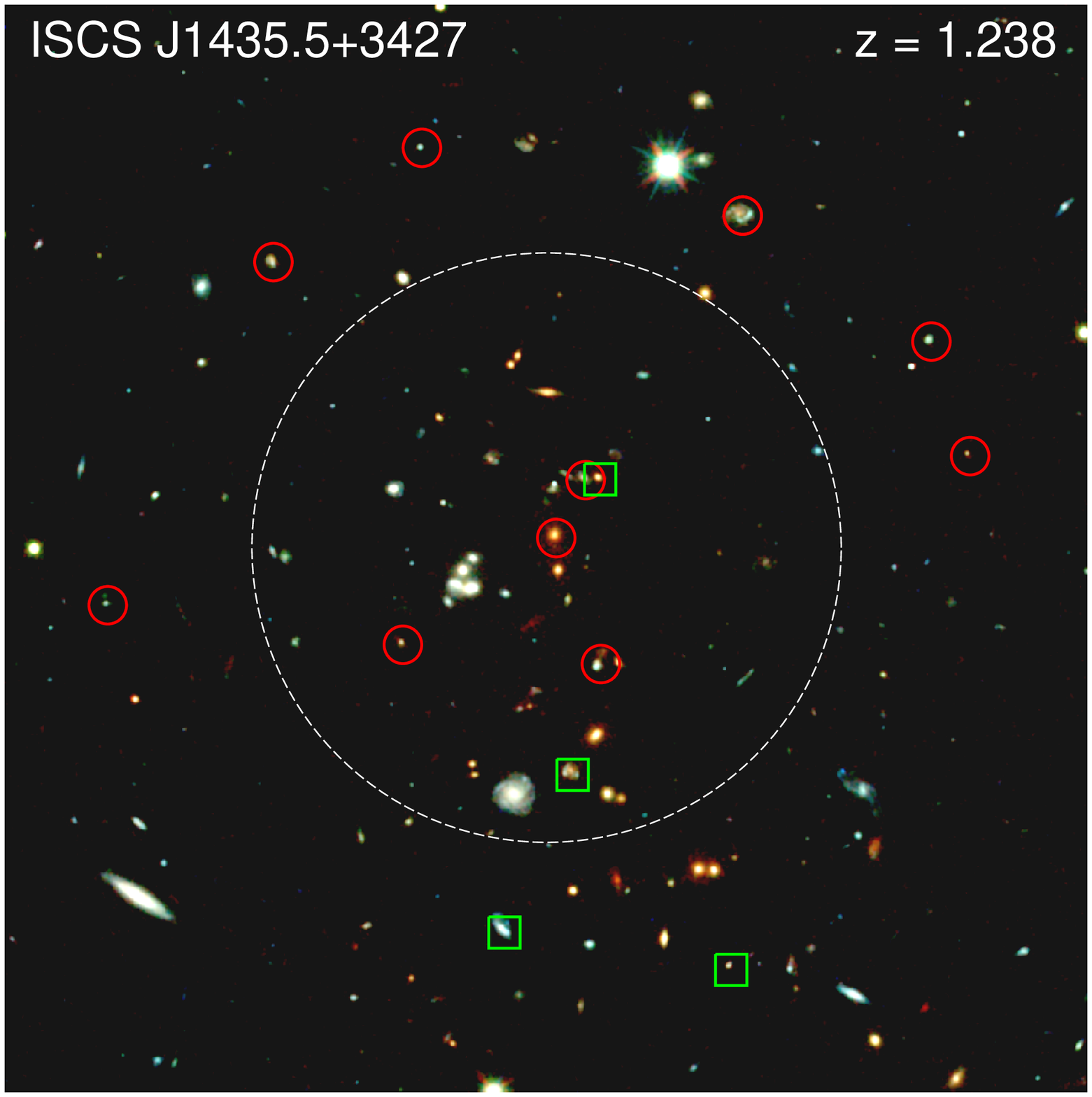}}
\fbox{\includegraphics[width=.45\textwidth]{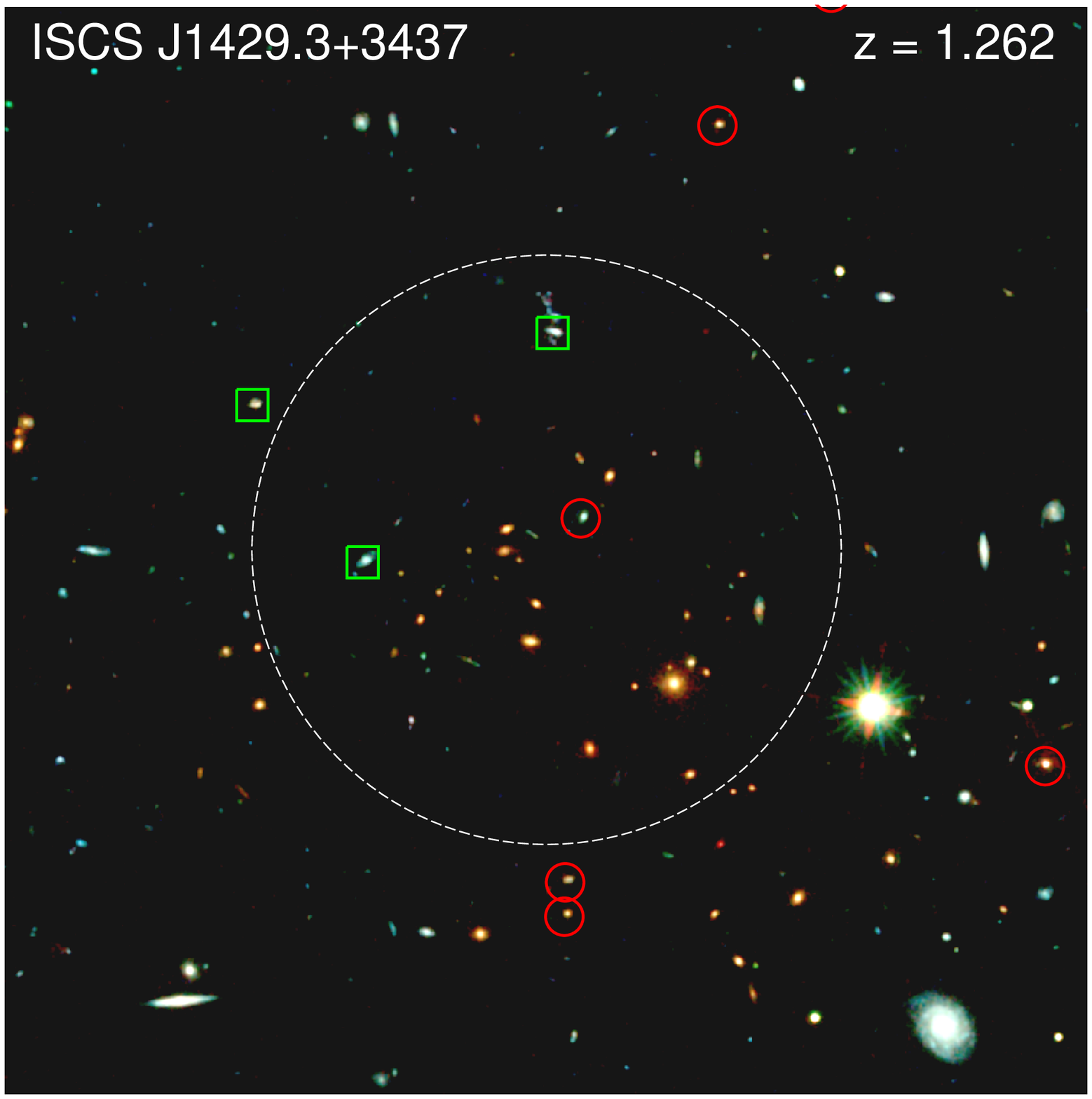}}
\fbox{\includegraphics[width=.45\textwidth]{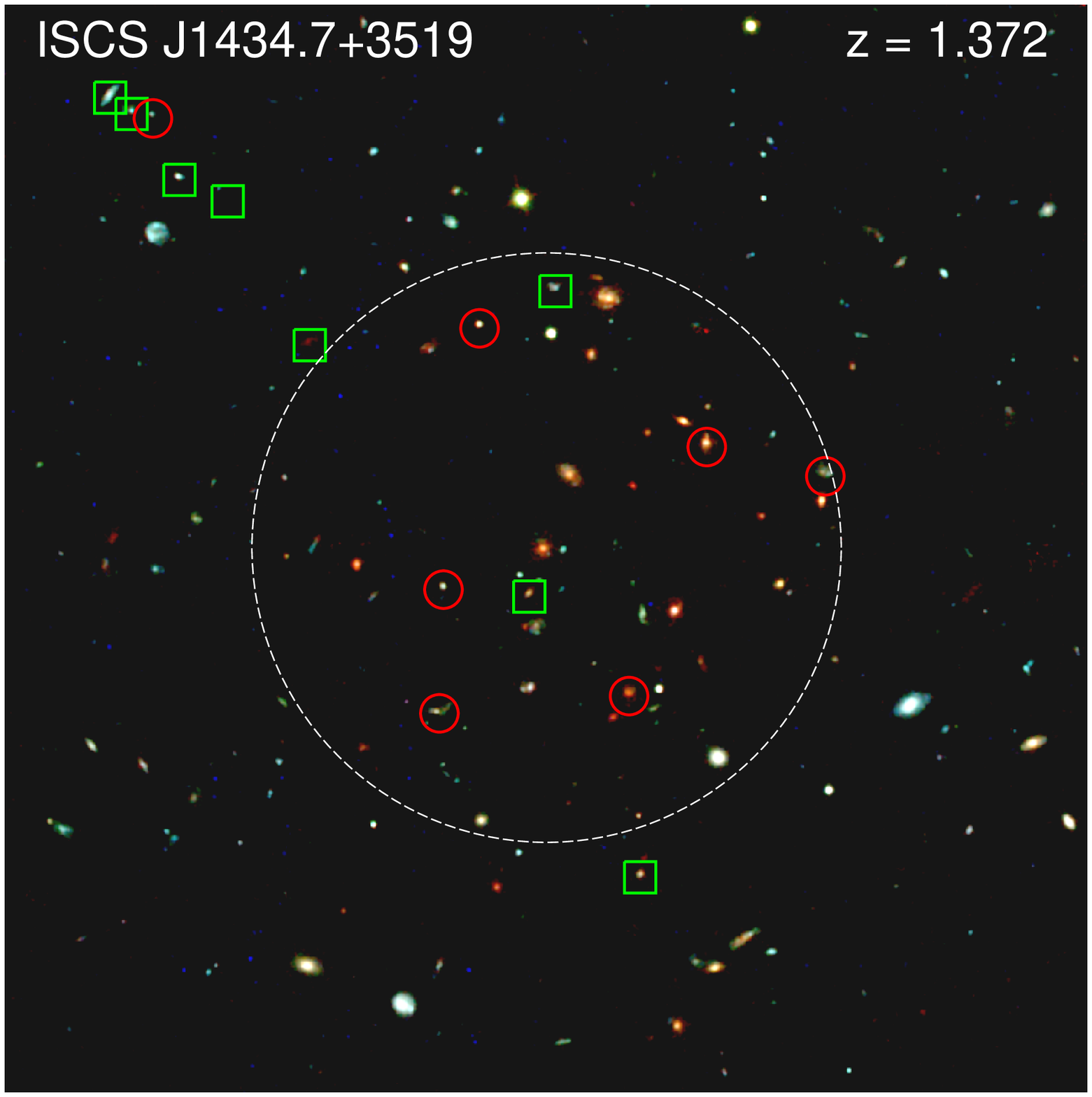}}
\fbox{\includegraphics[width=.45\textwidth]{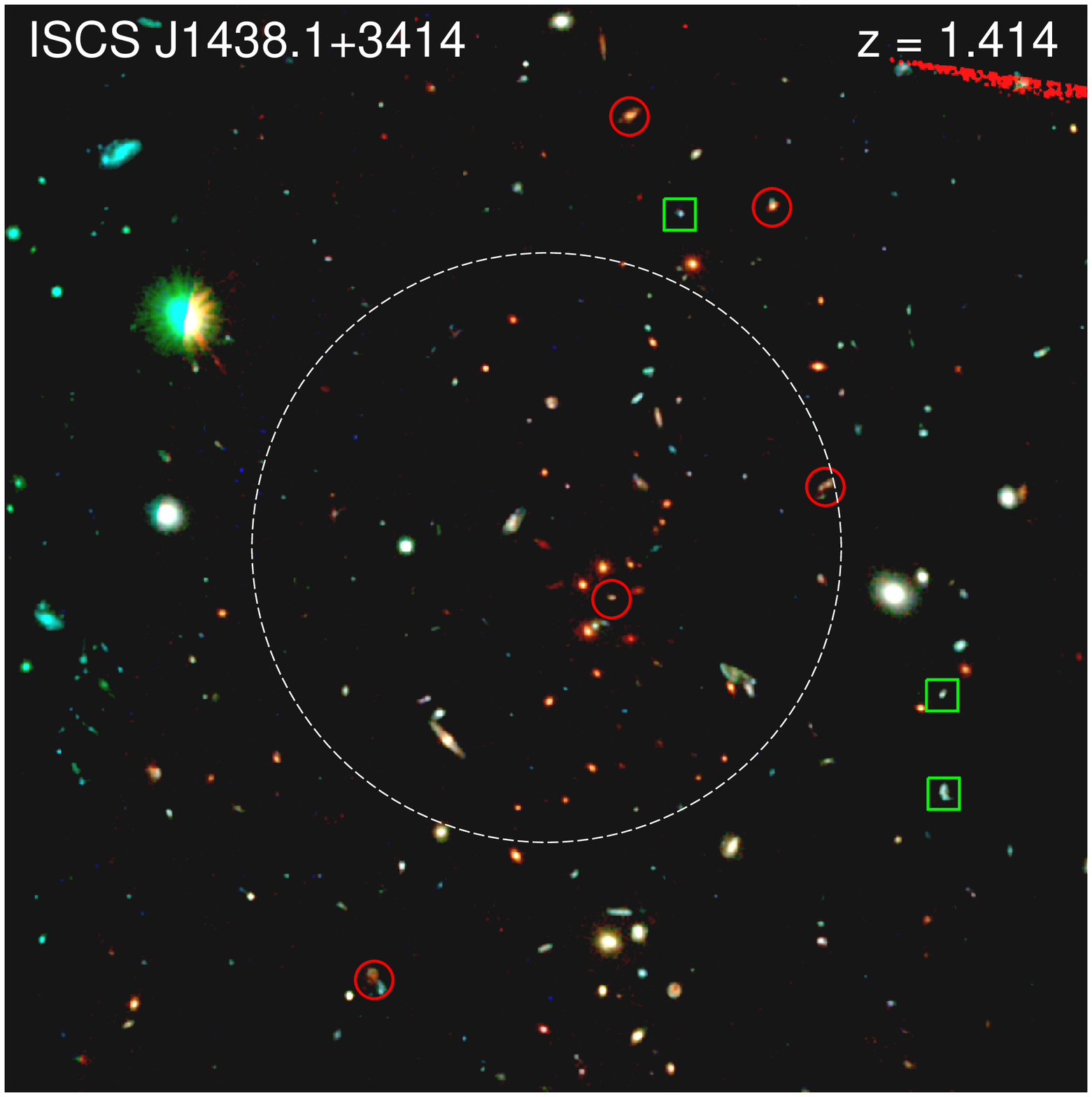}}
\centering
\caption{{\it HST} pseudo-color images (F775W+F850LP+F160W) for four of the clusters studied in this work.  The images are $\sim$90" x 90" on a side which is roughly 750kpc x 750kpc for these redshifts.  The images are centered on the clusters and the white dashed circle represents a 200 kpc radius, the size of the quenching radius discussed in \citet{bauer2011} and \citet{grutzbauch2012} for a massive galaxy cluster XMMU J2235.3-2557 at $z$ = 1.39 ($\sim$ 9$\times$10$^{14}$ M$_{\odot}$ yr$^{-1}$).  Red circles mark H$\alpha$ emitting cluster members while green squares signify H$\alpha$ emitting field galaxies.}\label{fig:f3}
\end{figure*}

\section{{\it HST} Data Reduction}
\label{sec:datared}

The data reduction process starts with the calibration of the raw images, both grism and direct. This was done automatically by the {\it HST} Data Archive (HDA) which runs CALWF3\footnote[9]{http://www.stsci.edu/hst/wfc3/documents/handbooks/} using the latest reference files.  The program wf3ir orchestrates the calibration process, which flags bad pixels, measures and subtracts the bias, corrects for non-linearity, flags saturated pixels, subtracts the dark image, calculates the flux conversion, converts data from counts to counts per second, flat fields the image, and calculates the gain conversion.  This process is the same for both the direct images and the grism images, with the exception of the flat fielding step.  The grism images are flat fielded at a later stage using the aXe\footnote[10]{http://axe.stsci.edu/} software and a master sky flat.  

 In slitless spectroscopy, a direct image is a necessary companion to the grism image in order to calibrate wavelength and properly identify and extract spectra.  The positions of objects detected in the direct image are used to establish the location of the corresponding spectra in the grism image.  Also, the size of the objects detected in the direct image are used to define the size of the box used for extraction.  It is therefore necessary to make a master catalog of sources detected in the direct image to be used later in the spectral extraction process.
 
The direct images were reduced using MultiDrizzle software (\citealp{fruchter2009}) and the resulting distortion-corrected, cosmic-ray rejected, coadded image was run through \textsc{SExtractor} (\citealp{bertin1996}) to produce a master catalog.  We use a detection threshold of 3.0$\sigma$ and detection minimum area of 6 pixels.  The catalog included all sources from the SExtractor extraction except for objects on the edges ($\pm$10 pixels).  The positions of the objects were then projected back to each individual direct image.  This is done because the 2D spectra are extracted from individual grism images and then coadded.

After the master catalog was created from the reduced direct image, we reduced the grism image using the program aXe (version 2.1).  The steps used to extract spectra are very similar to that found in WFC3 Grism Cookbook\footnote[11]{http://www.stsci.edu/hst/wfc3/analysis/grism\_obscookbook.html}.  The grism reduction process begins with the task AXEPREP which checks the units in the grism image and then subtracts a scaled master sky image to create a uniform sky background.  The master sky image\footnote[12]{http://www.stsci.edu/hst/wfc3/analysis/grism\_obs/\\calibrations/wfc3\_g141.html} used was the latest and best available at the time of the reduction.  This is important because a high S/N master sky image is required to extract spectra of faint objects.  

At this step, the individual grism images were ready for 2-D spectral extraction.  The extraction process was performed using the task AXECORE which defines the extraction geometry, flat-fields the region containing spectral information for each source, and determines the contamination from overlapping spectra.  The extraction geometry for our program is linked to the object's shape in the direct image in order to optimize the extraction of each spectrum.  We used a variable extraction width ($\pm$4 times the projected width of the source in the direction perpendicular to the spectrum trace) and an extraction direction in the direction of the dispersion with a tilt parallel to the orientation of the object (option 3 in \S 2.4 of \citealp{kummel2009}).  

Overlapping spectra are a significant issue in slitless spectroscopy.  When more than one spectrum contributes to the flux in a single pixel, we define that as contamination.  Contamination can occur in spatial or dispersion directions, and can come from other dispersed orders of objects that are not the target being extracted.  To estimate the contamination for each object we used a Gaussian emission model (\citealp{kummel2009}) which uses the broadband magnitude from the direct image and the size of the object to model a 2-D Gaussian emission spectrum centered on the central wavelength of the filter in the direct image.  This is done for all objects to create a contamination map.  The contamination map is extracted using the same geometry defined in the science extraction so that in a later step the contamination can be subtracted off in the 1-D spectrum space.

The extraction process is run on each individual grism image,  producing a 2-D extracted spectrum for each object in each image.  The 2-D extracted spectra for a common object are run through DRZPREP and AXEDRIZZLE to reject cosmic rays and coadd the spectra to produce a higher S/N 2-D spectrum (\citealp{kummel2004,kummel2005}).  The drizzle software is the standard software for combining {\it HST} images (\citealp{fruchter2009}) which properly handles weights and produces a deep 2-D grism spectrum for 1-D extraction.  We use an optimal extraction method, discussed in \citet{kummel2008} which employs a weighting scheme based on the Gaussian emission models discussed earlier.  The output of AXEDRIZZLE is a coadded 2-D spectrum as well as an optimally extracted 1-D spectrum that includes flux, error on the flux, and contamination in units of flux.

The final step in the extraction process is the creation of a webpage that combines the 2-D grism images with the 1-D extracted spectra in a visually useful format.  The program aXe2web\footnote[13]{http://axe.stsci.edu/axe/axe2web.html\#ref\_1} uses an input catalog and the aXe output files to create a webpage summary of the full reduction.  Each object is displayed on a separate row with the magnitude of the object, X and Y positions, the right ascension and declination, a direct image cutout, a grism image cutout, and a 1-D spectrum in counts and flux.  This webpage format of the spectral extractions provides an easy way to view the summary of the reductions for quality control as well as further science purposes such as emission line identification.           

\section{Measurements}

Using the webpage format of the grism reductions from aXe2web, emission lines in the 1-D extracted spectra were identified by eye and inspected in detail.  The strongest emission lines identified in the 1-D extracted spectra were assumed to be H$\alpha$, [O\textsc{iii}], or [O\textsc{ii}].  Other commonly detected emission lines include H$\beta$ and [S\textsc{ii}].  If only a single emission line was identified in a spectrum, then it was assumed to be H$\alpha$.  A redshift quality scale was used to quantify the robustness of the measurement.  A spectrum exhibiting a single feature was given a quality value of Q$=$C, while a spectrum showing two features consistent with the same redshift was assigned a quality value of Q$=$B, and a spectrum with three or more features indicating a single redshift was denoted with a quality value of Q$=$A.  Examples of all three quality redshifts are included in Figure 2.  Only robust redshifts, Q$=$A or Q$=$B, were included in the star formation rate analysis.

\begin{figure}[htp] 
\includegraphics[width=.85\textwidth]{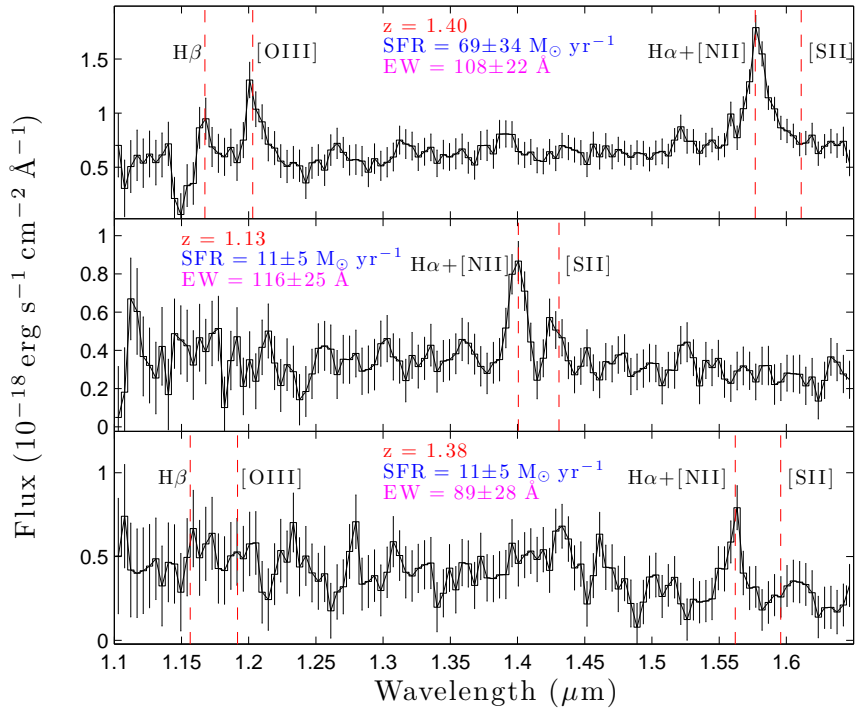}
\centering
\caption{Example WFC3 grism spectra of H$\alpha$ emitting cluster members.  The panels from top to bottom represent an example Q$=$A, Q$=$B, and Q$=$C redshift, respectively.  Typical emission lines are identified with red dashed vertical lines.  Also included in each panel is the redshift, star formation rate, and equivalent width from H$\alpha$ emission.}\label{fig:f3}
\end{figure}

Emission lines detected as H$\alpha$ were run through a custom program to measure both the line flux and equivalent width (EW).  A fourth order polynomial was fit to the continuum, excluding regions of emission.  A Gaussian (wavelength constrained to the identified peak $\pm$ 50\AA , width constrained to the spatial extent of the object $\pm$ 23\AA , and height unconstrained) was fit to the continuum-subtracted emission line to both measure the flux and the equivalent width (EW).  Errors for both the flux measurement and the EW were estimated from the 16$^{th}$ and 84$^{th}$ percentiles of 1000 realizations of the data assuming Gaussian errors.  Contamination from overlapping objects is subtracted off prior to the continuum fit and for $\sim$10\% of our sources the estimated contamination was $>$20\% in the continuum around H$\alpha$.  These sources are still included in our sample.  Exclusion of these sources does not effect our results.     

A finite width is used in the extraction of the 2D grism spectra, and we expect some flux to be lost due to this.  Employing the software aXeSim\footnote[14]{http://axe.stsci.edu/axesim/}, we used template emission line spectra with known H$\alpha$ and [N\textsc{ii}] fluxes to estimate the flux lost due to our finite extraction width.  A high resolution emission template was used with constant flux density, f$_\lambda$, and a constant strength of H$\alpha$+[N\textsc{ii}] emission relative to the continuum.  Twenty five different template spectra with magnitudes ranging from 19 to 24 AB which determined the continuum flux level and line flux of our template galaxy were used as input for aXeSim (all templates had the same H$\alpha$$+$[N\textsc{ii}] EW and H$\alpha$ flux ranged from $4\times10^{-17}$ ergs cm$^{-2}$ s$^{-1}$ to $4\times10^{-15}$ ergs cm$^{-2}$ s$^{-1}$).  The program aXeSim uses a master catalog and template spectra as input to create a simulated direct image and grism image from which spectra can be extracted.  The objects simulated were aligned in a vertical row with sizes and orientations that were representative of our galaxies, sampled directly from our catalogs, and all placed at $z=1$.  The simulated spectra were extracted using the same method described above and run through the line flux measurement program.  We found that 91$\pm$3$\%$ of the flux of H$\alpha$ was recovered (and this was constant across all H$\alpha$ input fluxes), with much of the loss due to the extraction aperture size and not the fitting method.  We apply a correction factor to our H$\alpha$ measurements to account for this flux loss in the extraction process.

\begin{figure}[htp] 
\includegraphics[width=.85\textwidth]{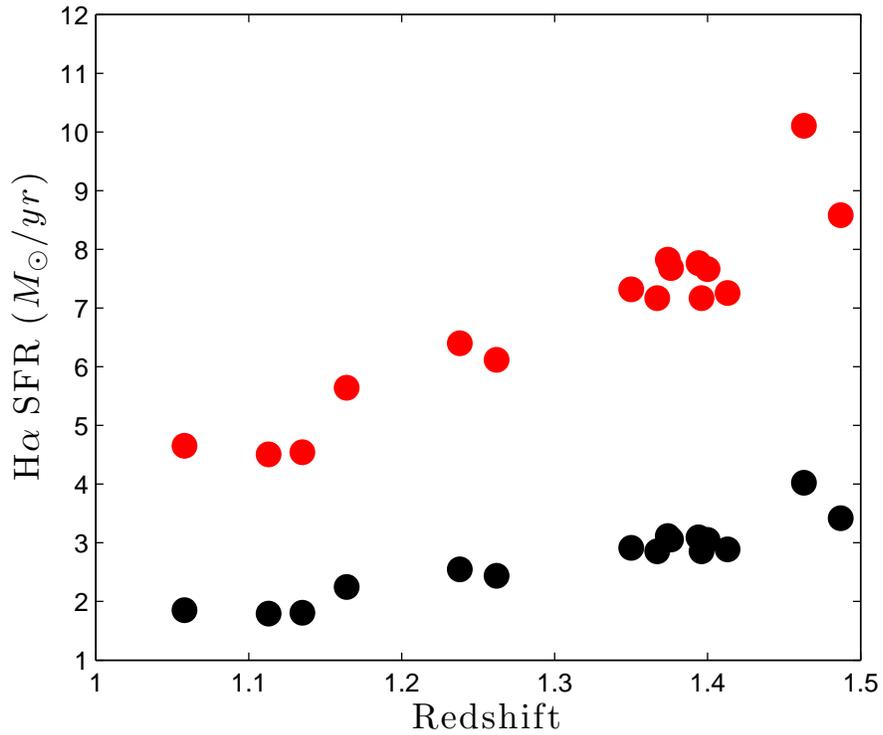}
\centering
\caption{SFR depth of our observations as a function of redshift.  The depths were determined from simulations.  The black points are the 50\% completeness limit (or recovery rate) of simulated sources whose redshifts were Q$=$A or Q$=$B extrapolated to each cluster given the total background level and G141 throughput curve.  The red points are the same as the black points but corrected for extinction using A$_{\rm H\alpha} = 1$.}\label{fig:f3}
\end{figure}

Also using the aXeSim software, we simulated grism images to estimate the depth of our observations.  We simulated images with the same exposure time, object positions, and magnitude distributions of some of our clusters.  All sources were given the same redshift of $z$ $=$ 1.  The simulated spectra were extracted using the same method as our observations and redshifts were identified and classified in a similar manner.  The 50\% completeness limit (or recovery rate) in flux, $\sim1 \times 10^{-16}$ ergs cm$^{-2}$ s$^{-1}$, was determined as the flux level at which 50\% of the simulated sources would have been included in our sample.  This was based on robust redshifts of our simulated spectra and was converted into a SFR as shown below.  The depth of our observations for different redshifts depends on the G141 wavelength throughput curve\footnote[15]{http://www.stsci.edu/hst/wfc3/documents/handbooks/currentIHB/c08\_slitless4.html} as H$\alpha$ falls at different wavelengths across the grism, and for different clusters the depth also depends on the background level of our grism exposures.  We extrapolated our simulation depth to each individual cluster (see Figure 3) by multiplying the 50\% completeness limit of the simulation by two separate factors: the square root of the background level normalized to the simulated background level and the throughput of the G141 grism at $\lambda = 6563 \times (1 + z_{\rm cluster})$ \AA\ normalized by the throughput of the G141 grism at 13126 \AA\ (H$\alpha$ at $z = 1$).      

\subsection{AGN Rejection}

H$\alpha$ emission can be due to high rates of star formation, AGN activity, or a combination of the two.  Since typical diagnostic emission lines (e.g., H$\alpha$ and [N\textsc{ii}]) are blended at the WFC3 grism resolution, we used X-ray observations from XBo\"{o}tes (\citealp{murray2005,kenter2005}) and deeper X-ray data centered on 13 of the clusters (\citealp{martini2012}) combined with the empirical mid-IR criteria from \citet{stern2005} to distinguish AGN from star-forming galaxies.  X-ray AGN were identified using a simple positional match in catalog space and a matching radius of 2$\arcsec$.  Also, objects matched to SDWFS catalogs with S/N $\geq$ 5 in all four IRAC bands that fell in the \citet{stern2005} AGN wedge were deemed AGN.  There were 27 sources (12 in the field and 15 in the clusters) satisfying either of these AGN criteria with 1.0 $< z <$ 1.5 and they were removed from this star formation analysis.  A more in-depth study of the AGN for these galaxy clusters was conducted by \citet{martini2012} and found evidence that the cluster AGN population has evolved more rapidly than the field population from $z \sim$1.5 to the present.

\subsection{Redshifts}

A total of 18 clusters, listed in Table 1, were observed with the WFC3 grism.  Redshifts were assigned for all emission-line objects, both cluster members and interloping field galaxies, and were determined solely from grism spectroscopy as the average redshift of multiple features (if present) or the redshift of just a single feature.  Only galaxies with robust redshifts, Q=A or Q=B, were used in the star formation analysis of this paper.  The resolution of the G141 grism allows a redshift identification to a precision of ${\sigma}_z = 0.01$.  These clusters have five to twenty or more total spectroscopic members from both this work and Keck spectroscopy (see \citealp{brodwin2013} for summary).  Due to the limited number of members and accuracy of the grism redshifts, we defined an H$\alpha$ emitting galaxy as a cluster member if $-0.03 < z - \langle z_{\rm cluster}\rangle  < 0.03$ (see Figure 4) while all other H$\alpha$ emitters with 1.0 $<$ $z$ $<$ 1.5 were considered field galaxies.  Note that below $z \sim 1.2$, galaxies are less likely to exhibit two strong emission features necessary for a robust redshift (see Figure 4).  This bias is not significant to our results.  If we restrict ourselves to $z > 1.2$ or include Q=C sources the general trends do not change (although they are at lower significance due to sample size or field/cluster confusion, respectively).   

\begin{figure}[htp] 
\includegraphics[width=.45\textwidth]{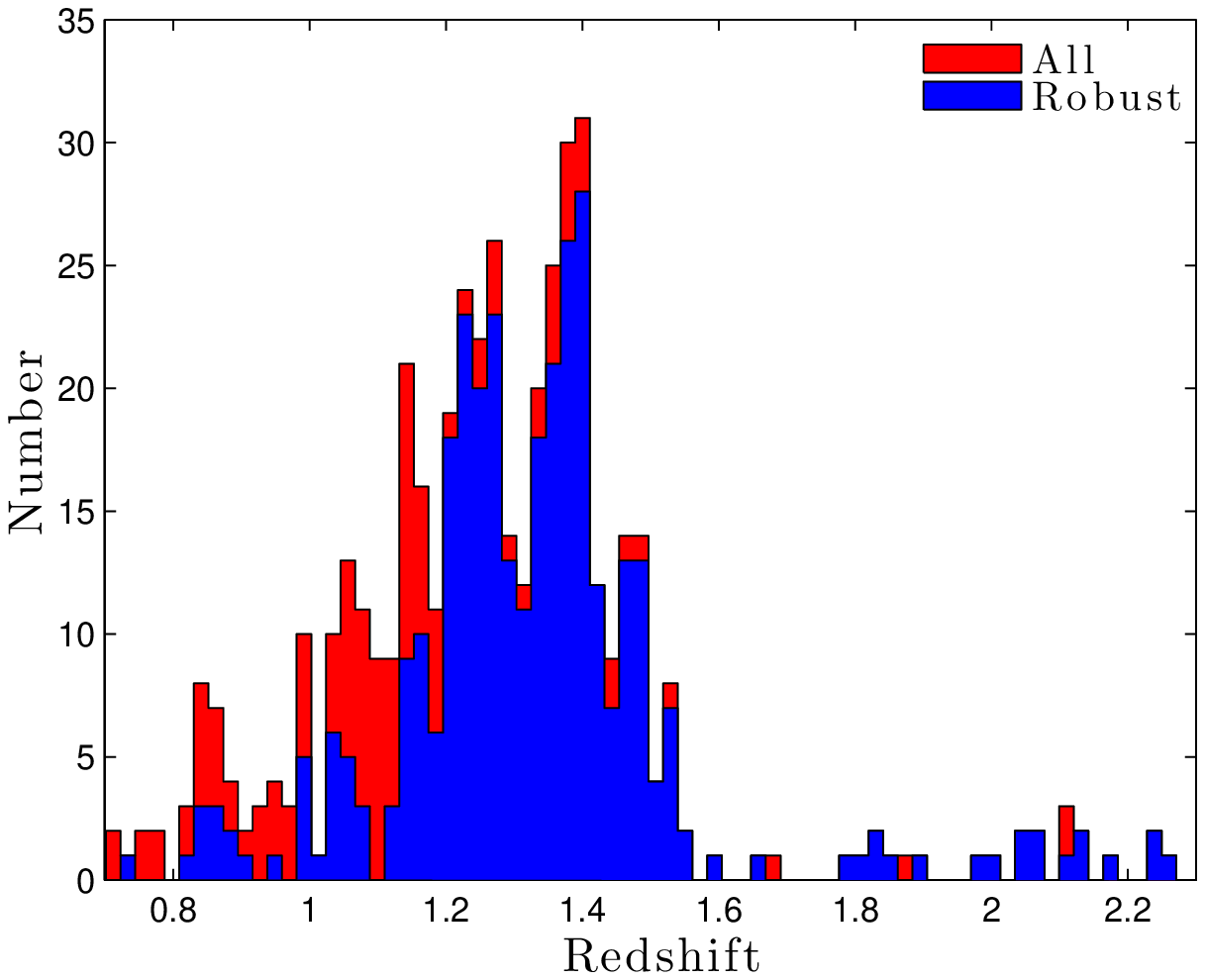}
\includegraphics[width=.45\textwidth]{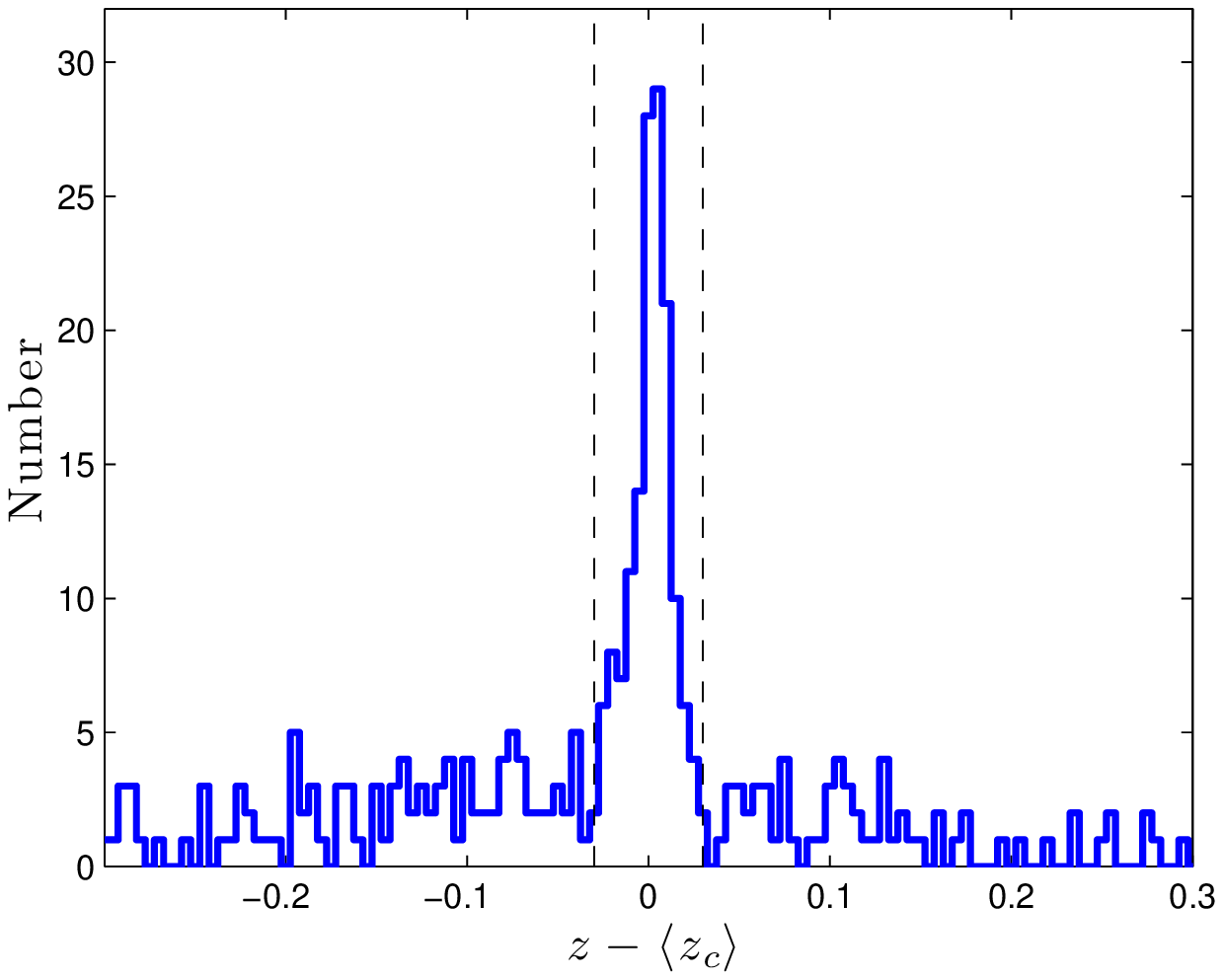}
\centering
\caption{{\it Left} -- Histogram of the redshift distribution for all identified emission line galaxies in red and robust redshifts in blue (Q$=$A or Q$=$B).  {\it Right} -- Histogram of the redshift distribution for all cluster fields shifted to the cluster redshift.  H$\alpha$ emitters enclosed by the vertical black dashed lines are defined as cluster members ($-0.03 < z - \langle z\rangle  < 0.03$) while all other H$\alpha$ emitters are considered part of the field population.}\label{fig:f3}
\end{figure}

H$\alpha$ emitters were identified for all 18 clusters although evidence for two of the clusters suggests that they are actually projected structures and not one bound configuration (ISCS J1429.2+3425 and ISCS J1427.9+3430), and are therefore not included in the following star formation analysis.  The confirmation of a new cluster is included in this work, ISCS J1437.0+3459 at $z$ $=$ 1.394.  The cluster includes six H$\alpha$ emitters and meets the criteria of a spectroscopically confirmed cluster defined in \citet{eisenhardt2008}, which holds that a cluster is confirmed if there are at least five cluster members within a radius of 2 Mpc whose spectroscopic redshifts match to within $\pm2000$($1+ \langle z_{\rm spec}\rangle$) km s$^{-1}$.    

A complete list of all redshifts from the grism observations are presented in Table 2.  

\subsection{Stellar Masses}

We estimate stellar masses for the H$\alpha$ emitters in our sample using iSEDfit (\citealp{moustakas2013}), a Bayesian spectral energy distribution (SED) fitting code that uses population synthesis models to infer the physical properties of a galaxy given its observed broadband SED.  We adopt the \citet{bruzual2003} population synthesis models, the \citet{chabrier2003} IMF, the \citet{calzetti2000} extinction curve, exponentially declining star formation histories with stochastic bursts of star formation superposed, and redshifts from our grism analysis.         

Our photometry, which reaches similar depths for all cluster fields, is derived using PyGFit (\citealp{mancone2013}).  PyGFit a python-based code that runs \textsc{GALFIT} (\citealp{peng2010}) on high-resolution images to fit a S\'{e}rsic galaxy model.  The model fit in the high-resolution image is then convolved with the point-spread function (PSF) of other bands to fit photometry.  We used the F160W filter as our high resolution image for galaxy model fits and measured photometry in B$_{W}$RIJHK+[3.6][4.5] for SED fitting.  Unrealistic model fits in the F160W imaging (S\'{e}rsic indices greater than 7.5 and/or effective radii greater than 20 kpc) were removed from this analysis.  This only effected five of the galaxies in our sample.  We also rejected galaxies with unconstrained stellar mass fits (errors greater than 0.4 dex or reduced ${\chi}^2$ fits greater than 2).  These issues arose due to lack of photometry redward of the 4000\AA\ break ($\sim$ 8\% of our sample), only a single or noisy photometric measurement redward of the 4000\AA\ break ($\sim$ 10\% of our sample), or poor fits due to an incorrect spectroscopic redshift identification (expected at a 5-10\% level for our robust redshifts).  The H$\alpha$ flux distribution of the sources rejected at this stage is similar to the final sample.   

Our final sample with good stellar masses includes 72 H$\alpha$ emitters in clusters and 76 H$\alpha$ emitters in the field.  For this sample, we are able to constrain the stellar mass quite well given our priors ($\sim$0.2 dex in logarithmic error), but because inferred physical properties from SED-fitting are dependent on the assumed stellar population synthesis models as well as the assumed priors, we use a stellar mass error of 0.3 dex for all masses.

\subsection{Star-formation Rates}

SFR is directly proportional to H$\alpha$ line luminosity (\citealp{kennicutt1998}) as this recombination line is sensitive to the most massive stars ($>$10 M$_{\odot}$) whose lifetimes are quite short ($<$20 Myr).  This relation (shown in Equation 1) assumes continuous star formation, Case B recombination at $T_e$ $=$ $10^4$ K, and is adjusted by a multiplicative factor of 0.64 to correct for a Chabrier initial mass function (IMF, \citealp{chabrier2003}).     
 \begin{equation}
{\rm SFR}_{\rm tot} = 5.0 \times 10^{-42} \times L_{\rm H\alpha} \times 10^{0.4 \times A_{\rm H\alpha}} = {\rm SFR}_{\rm H\alpha} \times 10^{0.4 \times A_{\rm H\alpha}}
\end{equation}
The last term in Equation 1 is used to correct for attentuation due to dust (A$_{\rm H\alpha}$).  We use the empirical relation of \citet{garn2010} and a Calzetti extinction law (\citealp{calzetti2000}), which relates dust attenuation to stellar mass (see Equation 2, where M=$\log{M_{*} / 10^{10} M_{\odot}}$).
\begin{equation}
A_{\rm H\alpha} = 0.91 + 0.77\ M + 0.11\ M^2 - 0.09\ M^3
\end{equation}
This relation was calculated with the same IMF used in our analysis and seems to hold out to $z \sim 1.5$ (\citealp{sobral2012}).

The H$\alpha$ line luminosity is calculated with the standard formula (see Equation 3), where $d_L$ is the luminosity distance for our adopted cosmology and $F_{\rm H\alpha+[N\textsc{ii}]}$ is what is measured.
\begin{equation}
L_{\rm H\alpha}=4 \pi {d_L}^2 F_{\rm H\alpha}
\end{equation}
\begin{equation}
F_{\rm H\alpha}= F_{\rm H\alpha+[N\textsc{ii}]} \times \frac{1}{1 + \frac{[N\textsc{ii}]}{H\alpha}}
\end{equation}
The spectral resolution of the G141 grism is 93 \AA\ (FWHM $\sim$ 2 pixels) which is sufficient to securely identify cluster members with a typical redshift accuracy of ${\sigma}_z$ $\approx$ 0.01 but blends H$\alpha$ and [N\textsc{ii}]$\lambda$6548,6584 emission.  The ratio of [N\textsc{ii}] to H$\alpha$ (specifically, [N\textsc{ii}]$\lambda$6584 / H$\alpha$) is commonly used as an indirect gas-phase metallicity indicator (e.g. \citealp{storchi1994,denicolo2002,maiolino2008}).  Recently, a tight relation, known as the fundamental metallicity relation (FMR), between stellar mass, SFR, and metallicity was observed both locally and at higher redshift (e.g. \citealp{mannucci2010,mannucci2011,belli2013,yuan2013}).  We can use this tight relation to infer the metallicities of our galaxies and convert those metallicities into an [N\textsc{ii}]$\lambda$6584 to H$\alpha$ ratio.  Equations 5 and 6 are from \citet{mannucci2011} (where M=$\log{M_{*} / 10^{10} M_{\odot}}$ and S=$\log{\rm SFR}$) who measured the FMR locally, but the relation seems to hold out to $z \sim 3$ (e.g. \citealp{belli2013}).  Stellar masses and SFRs used in \citet{mannucci2010,mannucci2011} were calculated with the same IMF and SFR indicator used in this analysis.    
\begin{equation}
12 + log (O / H) = 8.90 + 0.37 M - 0.14 S - 0.19  M^2 + 0.12  M  S - 0.54 S^2\ ({\rm for}\ 10^M - 0.32 S > 9.5)
\end{equation}
\begin{equation}
12 + log (O / H) = 8.93 + 0.51 \times (10^M - 0.32 S - 10)\ ({\rm for}\ 10^M - 0.32 S \leq 9.5)
\end{equation}
The metallicities used in \citet{mannucci2011} were empirical calibrated from \citet{maiolino2008}.  We can convert the metallicity from Equations 5 and 6 into a ratio of [N\textsc{ii}]$\lambda$6584 / H$\alpha$ using a relation found in \citet{maiolino2008} (shown in Equation 7 where T $=$ $(12 + [O / H]) - 8.69$).
\begin{equation}
\log{([N\textsc{ii}]\lambda6584 / \rm H\alpha)} = -.7732 + 1.2357\ T - 0.2811\ T^2 - .7201\ T^3 -.3330\ T^4
\end{equation}
Assuming a constant ratio of 3 to 1 for [N\textsc{ii}]$\lambda$6584 / [N\textsc{ii}]$\lambda$6548 and Equation 7, we can calculate [N\textsc{ii}] / H$\alpha$ and substitute back into Equation 4.  However, there is a complication; Equations 1 and 7 are coupled through 3, 4, 5, and 6.  To measure the SFR, one must know the ratio of [N\textsc{ii}] to H$\alpha$, but to know the ratio of [N\textsc{ii}] to H$\alpha$ one must know the SFR.  We solve these equations through iteration with an initial guess of [N\textsc{ii}] / H$\alpha$ $=$ 0.2, and solutions typically converge in less than five iterations.  

We assume the calculation above results in an error of 50\% for the total SFR and is dominated by the uncertainty in the extinction correction.            

\section{Results}

\subsection{Stacking}

A key aspect of this analysis is our correction of the H$\alpha$ unobscured SFR to the total star formation rate, assuming Equation 2.  The individual grism spectra are too noisy to accurately measure the Balmer decrement (specifically, the intensity ratio of H$\alpha$ to H$\beta$).  Instead to verify our approach, we performed a median stacking analysis of the 1-D extracted spectra to investigate the average reddening properties of the star forming galaxies.  By measuring the ratio of H$\alpha$ to H$\beta$ in a median stacked spectrum, we are able to estimate the extinction due to dust for the median object in the sample.

\begin{figure}[htp] 
\includegraphics[width=1\textwidth]{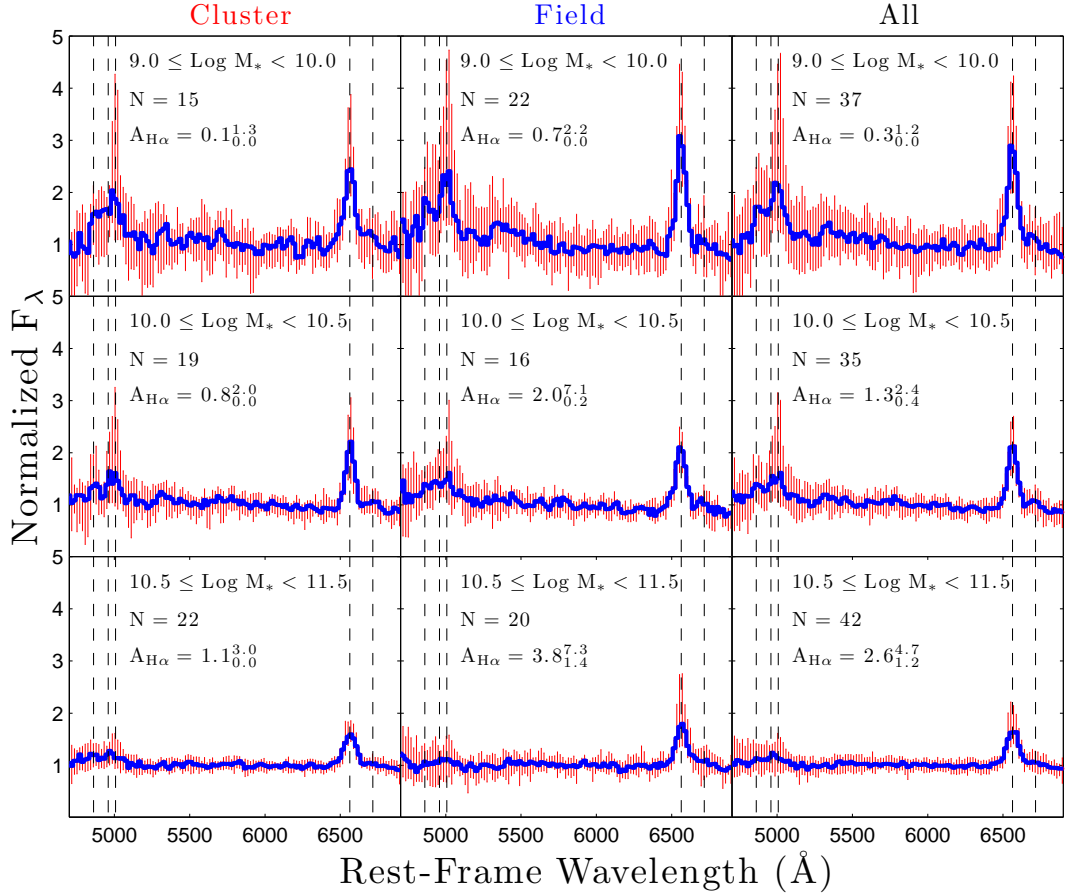}
\centering
\caption{Median stacked spectra for cluster, field, and all H$\alpha$ emitters (from left to right) at $z$ $>$ 1.24, binned by stellar mass (9.0 $\leq$ $\log$ M$_{*}$ $<$ 10.0, 10.0 $\leq$ $\log$ M$_{*}$ $<$ 10.5, and 10.5 $\leq$ $\log$ M$_{*}$ $<$ 11.5, from top to bottom).  The black dashed lines mark H$\beta$, [O\textsc{iii}]$\lambda$4959,5007, H$\alpha$+[N\textsc{ii}], and [S\textsc{ii}].  For each sample, the spectra were stacked 1000 times with bootstrap resampling and the median of those iterations is plotted in blue and the range of the 16$^{th}$ to 84$^{th}$ percentile is plotted with red error bars.  In each of the iterations, we subtracted an [N\textsc{ii}] contribution from H$\alpha$ drawn from a uniform distribution between 0.0 and 0.2 (95\% range of our individual galaxies calculated using Equations 1-7) to account for this uncertainty.  The number of galaxies in each sample are listed in the plots as well as the extinction for H$\alpha$ measured via the Balmer decrement (i.e. A$_{\rm H\alpha}$).  Note that all cluster members in this work are within a projected radius of 750 kpc as this is roughly the field of view of {\it HST} for $z \sim 1.2$.
 }\label{fig:f3}
\end{figure}  

Generating a median composite spectra requires re-binning the spectra to the rest frame, scaling the spectra, and stacking the spectra into a final composite using the median value at each re-binned wavelength (see \citealp{francis1991} for a detailed discussion of stacking).  We re-binned each spectrum to 18 \AA, approximately the pixel size of the observed spectra shifted to the rest frame, and scaled each spectrum by the median value of the continuum excluding regions of expected emission.  It is important to remember that the median spectrum preserves the relative fluxes of emission features.   

Assuming a \citet{calzetti2000} extinction law and an assumed ratio (H$\alpha$/H$\beta$) = 2.86 from case B recombination (\citealp{storey1995}),  we measured the extinction of the median stacked spectrum of star forming galaxies at $z$ $>$ 1.24, the redshift when H$\beta$ enters the wavelength coverage of the G141 grism.  We performed this stacking 1000 times using bootstrapping with replacement and found the extinction to be $E(B-V)$$=0.38$$^{\ 0.57}_{\ 0.21}$ (A$_{\rm H\alpha}$$=1.25$$^{\ 1.88}_{\ 0.68}$).  The upper and lower ranges were estimated with the 16$^{th}$ and 84$^{th}$ percentiles of the 1000 bootstrap realizations.  This is consistent with the calculated median value and range (16$^{th}$ and 84$^{th}$ percentiles), A$_{\rm H\alpha}$ $=1.09$$^{\ 1.48}_{\ 0.64}$, from Equation 2.  

We also fit the spectrum using a linear continuum model and eight Gaussians (H$\beta$, [O\textsc{iii}]$\lambda$4959,5007, H$\alpha$, [N\textsc{ii}]$\lambda$6548,6584, and [S\textsc{ii}]$\lambda$6717,6731).  We assumed fixed rest-frame wavelengths of the emission lines and further constrained the ratios of the [O\textsc{iii}] and [N\textsc{ii}] line doublets to be 1/3 as well as having the same full width at half maximum.  The ratio of [N\textsc{ii}] to H$\alpha$ is degenerate in the fit but ranges from 0 - 0.20 (95\% interval) with a median value of [N\textsc{ii}] / H$\alpha$  $=$ 0.08.  This ratio is consistent with the median value and range (95\% interval) of our calculation from Equation 7, [N\textsc{ii}] / H$\alpha$ $=0.04$$^{\ 0.20}_{\ 0.00}$ and measurements in the literature for lensed galaxies at 0.9 $< z <$ 1.5 ([N\textsc{ii}] / H$\alpha$ = 0.03-0.17; \citealp{wuyts2012}) and more massive galaxies at $z \sim 1.2$ (median [N\textsc{ii}] / H$\alpha$ = 0.18; \citealp{queyrel2012}).  Furthermore, the common emission line diagnostic of H$\beta$ / [O\textsc{iii}]$\lambda$5007 versus [N\textsc{ii}]$\lambda$6584 / H$\alpha$ (e.g. \citealp{baldwin1981}) suggests that the median stacked spectrum is that of a star-forming galaxy and not an AGN.         

Locally, extinction is positively correlated with stellar mass (e.g. \citealp{garn2010,zahid2013}), and this correlation is even seen at higher redshift ($z \sim1.47$; \citealp{sobral2012}).  We investigate if extinction is correlated in our sample of star-forming galaxies, as we assumed.  We split our sample into three different stellar mass bins (9.0 $\leq$ $\log$ M$_{*}$ $<$ 10.0, 10.0 $\leq$ $\log$ M$_{*}$ $<$ 10.5, and 10.5 $\leq$ $\log$ M$_{*}$ $<$ 11.5) for the cluster H$\alpha$ emitters, field H$\alpha$ emitters, and all H$\alpha$ emitters at $z$ $>$ 1.24 (the redshift when H$\beta$ enters the wavelength coverage, see Figure~5).  We find that there is indeed higher extinction for more massive star-forming galaxies, consistent with our assumed relation in Equation 2; however, because of low number statistics and noisy spectra we cannot constrain the correlation well.  Also, as seen in \citet{patel2011}, we find that for a fixed stellar mass, star-forming galaxies in the field have higher extinction on average than star-forming galaxies in clusters, albeit within the error bars.  If there is less dust per stellar mass in cluster star-forming galaxies than field star-forming galaxies, then there may also be less gas per stellar mass for new star formation. 

\subsection{Star Formation Analysis}

\begin{figure}[htp] 
\includegraphics[width=.85\textwidth]{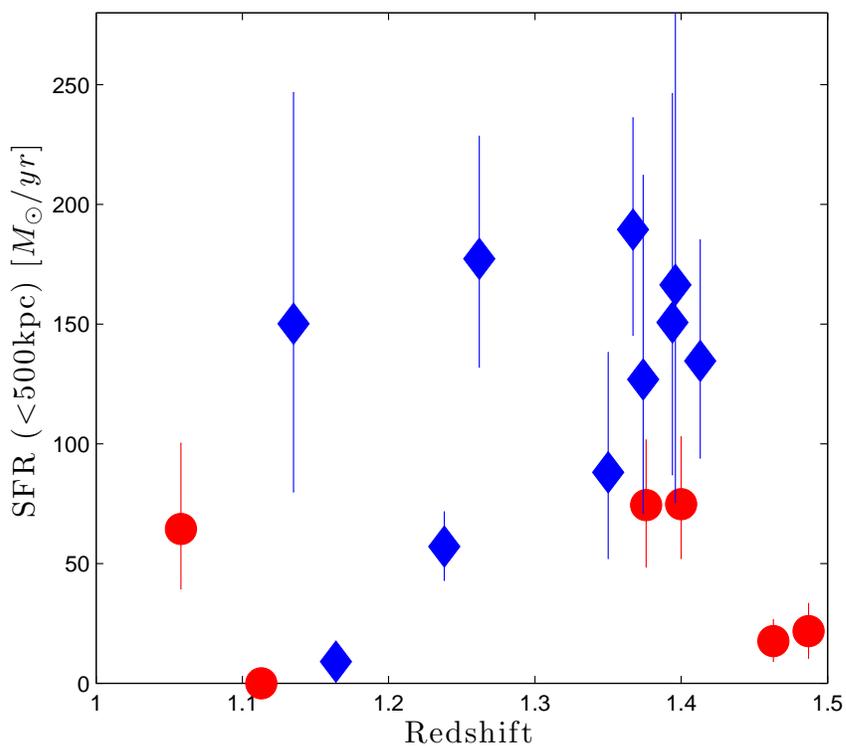}
\centering
\caption{The sum of the total (extinction-corrected) SFR within 500 kpc of our $z$ $>$ 1 clusters for galaxies with SFR$_{\rm H\alpha}$ $>$ 4 M$_{\odot}$ yr$^{-1}$ (50\% completeness limit at $z = 1.5$ for uncorrected H$\alpha$ SFR, see Figure 3).  Blue diamonds indicate clusters that have star formation within the inner 200 kpc, while red circles are those that do not.  There is a wide spread in total SFR within a 500 kpc radius and little evolution evident with redshift given the spread.  The error bars were estimated with the 16$^{th}$ and 84$^{th}$ percentiles of the 1000 bootstrap realizations.}\label{fig:f3}
\end{figure}

Figure 6 shows the sum of the total H$\alpha$ SFR (restricted to galaxies above the 50\% completeness limit at $z = 1.5$ from Figure 3, SFR$_{\rm H\alpha}$ $>$ 4 M$_{\odot}$ yr$^{-1}$) within a projected radius of 500 kpc (roughly the field of view of our observations) for each of the 16 clusters in this work.  There is a large scatter from cluster to cluster with a range of enclosed star formation rates from $0 - 200$ M$_{\odot}$ yr$^{-1}$ and little evidence for redshift evolution given the spread.   For a subset of our clusters with strong lensing cluster masses (six clusters with M$_{200}$ $\sim$ $(2.5-5) \times 10^{14} M_{\odot}$; \citealp{jee2011}), we found that the scatter in the enclosed SFR from cluster to cluster remains when normalized by cluster mass.  A larger sample with a mass and SFR measurement at the same enclosed radius is needed to investigate this further.

\citet{bauer2011} and \citet{grutzbauch2012} studied the H$\alpha$ SFRs in the galaxy cluster XMMU J2235.3-2557 at $z$ = 1.39 ($\sim$ 9$\times$10$^{14}$ M$_{\odot}$) and found a lack of star formation in the inner 200 kpc region, which was denoted as the quenching radius.  The majority of our clusters (10 of 16) have significant levels of unobscured star formation within a 200 kpc radius.  The six clusters that do not show evidence of star formation within a 200 kpc radius cover the entire redshift range of our sample (1.05 $<$ $z$ $<$ 1.49).  Our definition of a cluster member ($-0.03 < z - \langle z_{\rm cluster}\rangle  < 0.03$) allows for a large volume along the line of sight and may lead to projection effects.  When we used a more restrictive cut on a cluster member (spectroscopic redshifts match to within $\pm2000$($1+ \langle z_{\rm cluster}\rangle$) km s$^{-1}$) we still found 9 of 16 clusters have significant levels of unobscured star formation within a 200 kpc radius.

\begin{figure}[htp] 
\includegraphics[width=1\textwidth]{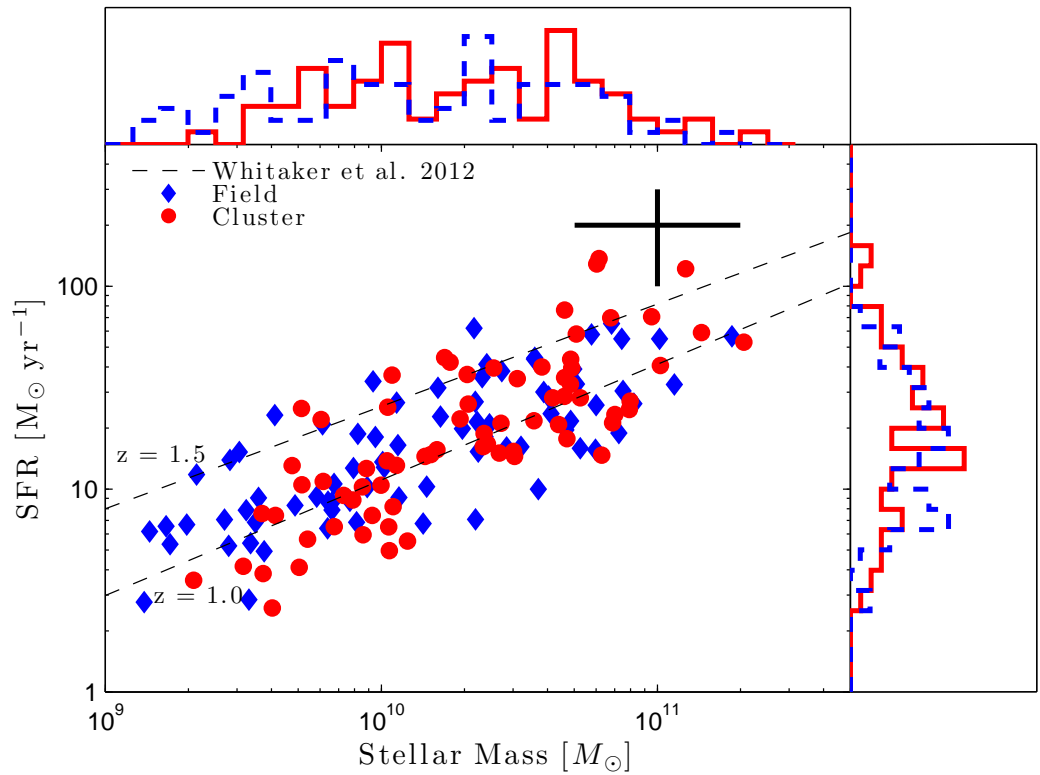}
\centering
\caption{SFR versus stellar mass.  The ``main sequence'' of star-forming galaxies at $z$ $=$ 1.0 and $z$ $=$ 1.5 is plotted as two dashed black lines (\citealp{whitaker2012}).  Normalized histograms of stellar mass and SFR for both the field (blue) and cluster (red) populations are plotted at the top and right, respectively.}\label{fig:f3}
\end{figure}

Figure 7 plots SFR versus stellar mass for both cluster and field star-forming galaxies.  We assumed an inherent correlation between stellar mass and SFR in Equations 1 and 2, which is a reasonable assumption since SFR and stellar mass form a ``main sequence'' of normal star-forming galaxies (\citealp{Noeske2007}).  At 1.0 $<$ $z$ $<$ 1.5, the distribution of star-forming galaxies in clusters are similar to that of the field and lie on the SFR-mass relation found in the literature (\citealp{whitaker2012}), confirming that our extinction correction for total SFR is statistically consistent with other works.    

\begin{figure}[htp] 
\includegraphics[width=.85\textwidth]{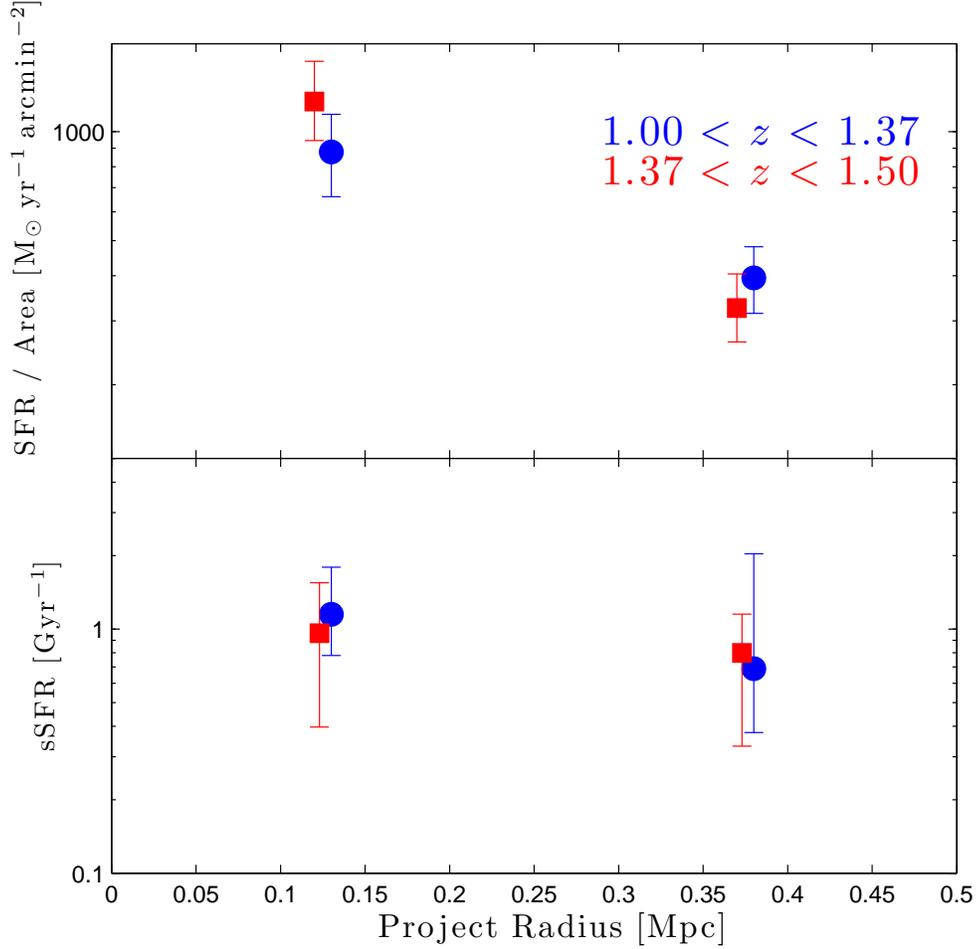}
\centering
\caption{({\it top panel}): SFR density versus cluster centric radius in projection.  Blue circles are galaxies in clusters at 1.00 $<$ $z$ $<$ 1.37 and red squares are galaxies in clusters at 1.37 $<$ $z$ $<$ 1.50, each binned by projected radius and slightly offset for clarity.  The error bars were estimated with the 16$^{th}$ and 84$^{th}$ percentiles of 1000 bootstrap realizations.  ({\it bottom panel}): Median sSFR versus cluster centric radius in projection.  The symbols are the same as the top panel.  There is no evidence for a decline in sSFR for galaxies closer in projection towards the cluster center.  The error bars were estimated with the 16$^{th}$ and 84$^{th}$ percentiles of 1000 bootstrap realizations.}\label{fig:f3}
\end{figure}

We also investigated the SFRs of cluster galaxies with respect to their cluster centric radius.  The top panel of Figure 8 displays the SFR density of cluster galaxies at two different redshifts (1.00 $<$ $z$ $<$ 1.37 and 1.37 $<$ $z$ $<$ 1.50 which splits the cluster sample in half).  There is a significant rise in the SFR density from the outer core to the inner core; however, there is no apparent redshift evolution in the relation.  The calculation for SFR density was corrected on a cluster to cluster basis for the grism FoV which typically was $\sim$1.1 Mpc $\times$ 0.9 Mpc, but varied as a function of redshift and for some clusters there were small offsets between the center of the observation and the center of the cluster.  Note that the stellar masses of the inner core are higher than those of the outer core and may be the cause of the trend.  Plotted in the bottom panel of Figure 8 is the median sSFR for the same two redshift bins as a function of cluster centric radius.  The median sSFR for both redshifts is roughly an order of magnitude higher than local star-forming galaxies in clusters (\citealp{wetzel2012}).  We found a flat trend in the median sSFR as a function of radius similar to what was found in Figure 10 of \citet{muzzin2012} at 0.85 $< z <$ 1.20.     

\citet{brodwin2013} also studied the (s)SFRs of this sample but using the 24$\mu$m luminous galaxies, more closely associated with obscured star formation and timescales on the order of $\sim$Gyr.  They found a similar rise in SFR-density with smaller radii, but they observed a redshift evolution which was most significant for clusters above $z$ $>$ 1.37.  Their observations went out to $\sim$2$\times$r$_{\rm virial}$, included photometrically selected sources (photo-z's), and were limited to starbursting galaxies ($\sim$50 M$_{\odot}$ yr$^{-1}$) with $\log_{10}$M$_{*}$ $>$ 10.1 (our sample only contains nine galaxies above these two limits).    

\subsection{Equivalent Widths}

\begin{figure}[htp] 
\includegraphics[width=.85\textwidth]{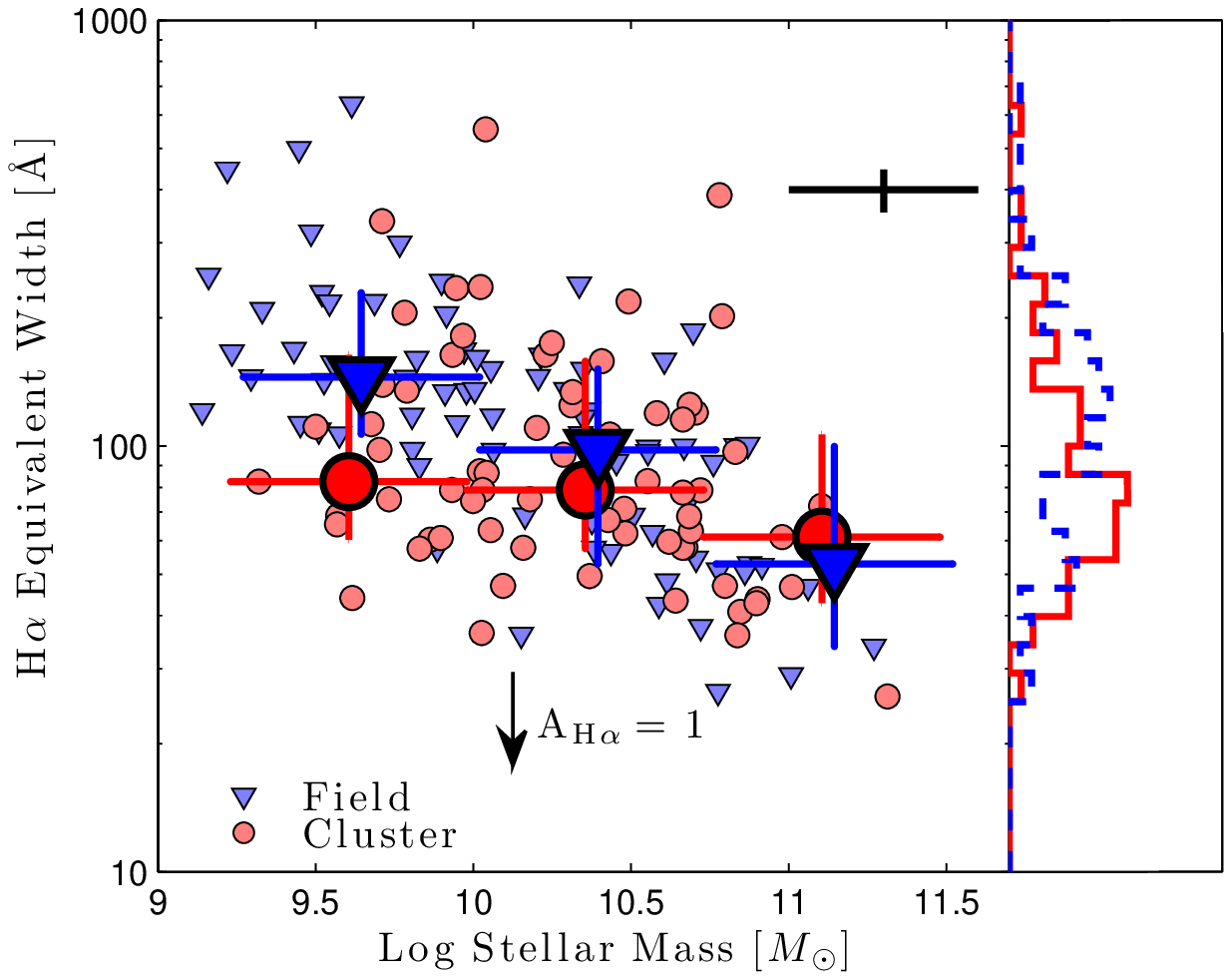}
\centering
\caption{Rest-frame H$\alpha$ equivalent width versus stellar mass.  The EW serves as a good proxy for specific star formation rate and shows the common relation that galaxies of higher stellar mass have on average lower EWs.  The blue triangle symbols are field galaxies and the red circles are cluster members.  The larger blue triangles and red circles are the median values of the field and cluster sample, respectively, in three different mass bins (9.25 $\leq$ $\log$ M$_{*}$ $<$ 10.0, 10.0 $\leq$ $\log$ M$_{*}$ $<$ 10.75, and 10.75 $\leq$ $\log$ M$_{*}$ $<$ 11.5).  The median values are offset slightly for clarity.  The error bars in vertical direction represent the widths of the EW distribution in each bin and were estimated with the 16$^{th}$ and 84$^{th}$ percentiles of 1000 bootstrap realizations.  The error bars in the median value itself are much smaller and the difference between the two median values in the lowest mass bin is at the $\sim$2.5$\sigma$ level. In each realization, we allowed the stellar masses to vary according to a Gaussian distribution with the mean value as the measured stellar mass and 0.3 dex as the standard deviation.  This gives a realistic estimate of the sample variation due to binning.  Also, a black arrow is plotted to show the effect of differential extinction for A$_{\rm H\alpha}$ $=$ 1 assuming a \citet{calzetti2001} relation (A$_{\rm 6563}$ $=$ 0.44 $\times$ A$_{\rm H\alpha}$).  On the right is a normalized histogram of the EW distributions for the field (dashed blue) and cluster (solid red) samples.  The Kolmogorov-Smirnov test reveals that there is only a 0.2\% chance that the two distributions are drawn from the same underlying parent population with the field galaxies exhibiting higher EWs on average.}\label{fig:f3}
\end{figure}

The H$\alpha$ EW (rest-frame) is proportional to the mass to light ratio at 6563\AA ~times the specific star formation rate (sSFR) and the differential extinction between the gas and stars (see Equation 8 where $F_{\rm 6563}$ is the flux density at 6563\AA , $L_{\rm 6563,cor}$ is the extinction-corrected luminosity density at 6563\AA , and A$_{\rm 6563}$ is the extinction from dust for the stellar continuum).  The last term in Equation 8 was included because some studies have found that the extinction due to dust experienced by continuum star light is less than that of nebular emission lines from gas (\citealp{calzetti2001}, A$_{\rm 6563}$ $=$ 0.44 $\times$ A$_{\rm H\alpha}$).  For simplicity, we assumed  A$_{\rm 6563}$ $=$ A$_{\rm H\alpha}$, and discuss later how a differential extinction would affect our EW measurements.
\begin{equation}
EW_{\rm H\alpha} = \frac{F_{\rm H\alpha}}{F_{\rm 6563}} = \frac{L_{\rm H\alpha}}{L_{\rm 6563}} \propto \frac{{\rm SFR}_{\rm H\alpha}}{L_{\rm 6563}} \propto \frac{{\rm SFR}_{\rm H\alpha}}{{\rm M}_{*}}\ \frac{{\rm M}_{*}}{L_{\rm 6563}} \propto \frac{{\rm SFR}_{\rm tot}}{{\rm M}_{*}}\ \frac{{\rm M}_{*}}{L_{\rm 6563, cor}}\ 10^{-0.4\ (A_{\rm H\alpha} - A_{\rm 6563})} 
\end{equation}

Plotted in Figure 9 is the EW versus stellar mass for both field and cluster star-forming galaxies.  Field star-forming galaxies, on average, have higher EWs than those in the cluster environment, and this is most noticeable for lower stellar masses ($\log$ M$_{*}$ $<$ 10.0).  According to the non-parametric Kolmogorov-Smirnov test, there is only a 0.2\% chance that the two distributions are drawn from the same underlying parent population (the Anderson-Darling test shows there is only a 1\% chance that the two distributions are the same).  This result holds if you include star-forming galaxies without ``good'' stellar mass estimates (see \S4.3).  There is no systematic difference in the selection of field versus cluster galaxies as they show the same magnitude distribution (F160W), were selected at the same flux limits, and were drawn from the same set of observations.  The sources without ``good'' stellar masses ($\sim$25\%) cover a wide range of F160W magnitudes for both the field and the cluster and are not relegated to the faint end.

Starting from Equation 8, it is easy to see that EW is sensitive to differential extinction as well as the calculated [N\textsc{ii}] to H$\alpha$ ratio as it enters through the calculation of $F_{\rm H\alpha}$.  For the two distributions to be the same, cluster star-forming galaxies would have to have more extinction at a given stellar mass than the field population.  From our stacking analysis in Figure 5, we see that this is not the case, and in fact, the opposite seems to be more likely.  To verify that the calculated ratio of [N\textsc{ii}] to H$\alpha$ does not have a large effect on the EW measurement, we performed the same analysis for a variety of assumed constant ratios, [N\textsc{ii}] / H$\alpha$ = 0.05, 0.10, and 0.20, and the results were the same.  The difference in EW distributions seems to be robust and not a systematic effect.

To investigate the difference between the two EW distributions further, it is informative to look at the right side of Equation 8.  The EW depends on the sSFR, mass to light ratio, and differential extinction.  As discussed earlier, accounting for differential extinction (emission from gas is attenuated more than star light) would only make the two distributions more disparate and does not explain why they are offset.  Since EW is proportional to sSFR, it might be expected that for the lowest mass bin in Figure 9 (9.25 $\leq$ $\log$ M$_{*}$ $<$ 10.0) the distribution of sSFR for the field would be higher than in the cluster environment.  This is seen in Figure 10, which plots the distributions for cluster (red) and field (blue) star-forming galaxies in stellar mass bins (from top to bottom) for SFR, sSFR, and EW (from left to right).  The difference between the field and cluster distribution is not as dramatic for sSFR as it is for EW.  The remainder of the difference may be attributed to different mass to light ratios for the field population compared to the cluster population.  This would reflect different star formation histories for the field and cluster galaxies at fixed stellar mass.     

\begin{figure}[htp] 
\includegraphics[width=.9\textwidth]{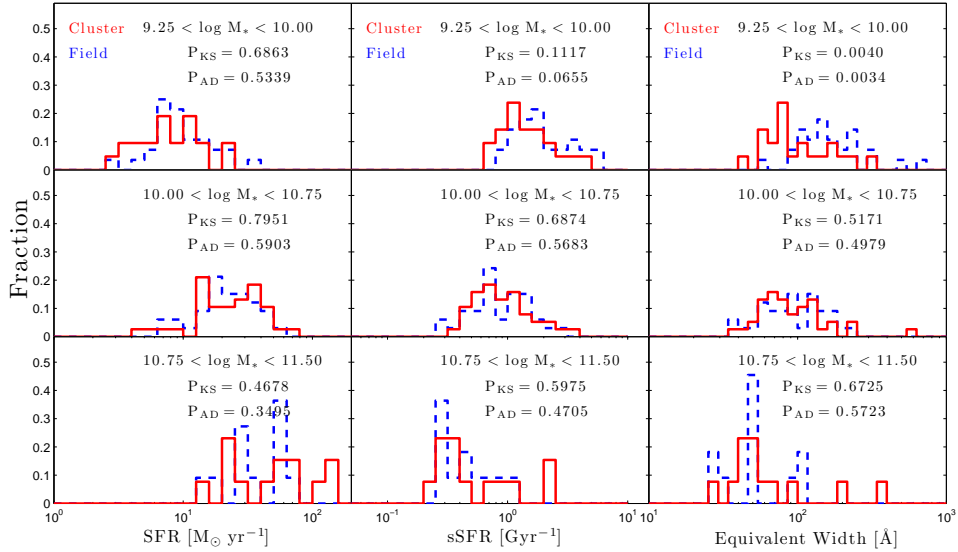}
\centering
\caption{Plotted are the histograms of the distributions for cluster (red) and field (blue) star-forming galaxies in stellar mass bins (from top to bottom) for SFR, sSFR, and EW (from left to right).  Also shown in each panel are the probabilities that the two distributions are drawn from the same parent population for the Kolmogorov-Smirnov test (P$_{\rm KS}$) and the Anderson-Darling test (P$_{\rm AD}$).  We ran both tests on 1000 bootstrap realizations of the data (allowing stellar mass to vary using a Gaussian distribution with a mean from the measured stellar mass and a standard deviation of 0.3 dex) and the median value of the test is quoted in each panel.}\label{fig:f3}
\end{figure}

\section{Conclusions}

Using the {\it HST}/WFC3 grism, we observed H$\alpha$ emission in the core of 16 M$_{200}$ $\sim$ ($1-5$) $\times$ 10$^{14}$ M$_{\odot}$ galaxy clusters.  The observations allowed us to identify cluster members and field galaxies at 1.0 $<$ $z$ $<$ 1.5 in a consistent way with identical selection methods.  Using a suite of multi-wavelength data, high-resolution imaging, and grism spectroscopy, we compared the average extinctions, SFRs, and EWs of star-forming galaxies as a function of stellar mass, redshift, and environment.  Our key findings are:

\begin{enumerate}
\item We find tentative evidence that extinction is a function of stellar mass for star-forming galaxies in both the cluster and the field environment with higher extinction values on average for the field than the cluster.  A larger sample size is needed to confirm this. 
\item There is a large scatter in the SFRs of the cluster cores ($<$ 500 kpc).  Also, many of the clusters (10 of 16) have high levels of star formation in the inner core ($<$ 200 kpc) which other studies have suggested to be a quenching radius for more massive systems (\citealp{grutzbauch2012}).
\item The H$\alpha$ EWs of the field star-forming galaxies are higher than those in clusters for $\log$ M$_{*}$ $\lesssim$ 10.  The suppression of EW in the cluster environment suggests that environmental effects are still apparent to at least $z$ $=$ 1.5.
\end{enumerate} 

AHG acknowledges support from the National Science Foundation through grant AST-0708490. This work is based in part on observations made with
the {\it Spitzer Space Telescope}, which is operated by the Jet Propulsion Laboratory, California Institute of Technology under a contract with NASA. Support for this work was provided by NASA through an award issued by JPL/Caltech. Support for {\it HST} programs 10496, 11002, 11597, and 11663 were provided by NASA through a grant from the Space Telescope Science Institute, which is operated by the Association of Universities for Research in Astronomy, Inc., under NASA contract NAS 5-26555.  This work makes use of image data from the NOAO Deep WideÐField Survey (NDWFS) as distributed by the NOAO Science Archive. NOAO is operated by the Association of Universities for Research in Astronomy (AURA), Inc., under a cooperative agreement with the National Science Foundation. 

We thank Matt Ashby for creating the IRAC catalogs for SDWFS, Buell Jannuzi for his work on the NDWFS, Michael Brown for combining the NDWFS with SDWFS catalogs, and Steve Murray and the XBo\"{o}tes team for obtaining the {\it Chandra} data in the Bo\"{o}tes field.  This paper would not have been possible without the efforts of the support staffs of the {\it Spitzer Space Telescope}, {\it Hubble Space Telescope}, and {\it Chandra X-ray Observatory}.  The work by SAS at LLNL was performed under the auspices of the U. S. Department of Energy under Contract No. W- 7405-ENG-48.    

\bibliography{halpha_sfr}

\begin{deluxetable}{cllcccl}
\centering
\tabletypesize{\scriptsize}
\tablecaption{ISCS WFC3 Grism High-redshift cluster sample}
\tablewidth{0pt}
\tablehead{
 \colhead{ID} & \colhead{R.A.} & \colhead{Dec.} &  \colhead{$<z_{\rm sp}>$} & \colhead{N$_{\rm H\alpha}$\footnotemark[1]} &\colhead{N$_{\rm spec}$} & \colhead{Reference} \\
\colhead{---} & \colhead{(J2000)} & \colhead{(J2000)} &  \colhead{---} & \colhead{---} & \colhead{---}
 }
 \startdata

ISCS J1429.2+3357 & 14:29:15.16 & 33:57:08.5 & 1.058 & 1 & 8 & \citealp{eisenhardt2008} \\
ISCS J1432.4+3332 & 14:32:29.18 & 33:32:36.0 & 1.113 & 2 & 26 & \citealp{elston2006,eisenhardt2008} \\
ISCS J1426.1+3403 & 14:26:29.51 & 34:03:41.1 & 1.135 & 2 & 12 & \citealp{eisenhardt2008} \\
ISCS J1429.2+3425 & 14:29:15.16 & 34:25:46.4 & 1.161/1.203 & 0 & 6/6 & \citealp{brodwin2013}; This Work \\
ISCS J1426.5+3339 & 14:26:30.42 & 33:39:33.2 & 1.164 & 5 & 14 & \citealp{eisenhardt2008} \\
ISCS J1427.9+3430 & 14:27:54.88 & 34:30:16.3 & 1.235 & 0 & 4 & \citealp{brodwin2013}; This Work \\
ISCS J1434.5+3427 & 14:34:30.44 & 34:27:12.3 & 1.238 & 6 & 19 & \citealp{brodwin2006,elston2006} \\
ISCS J1429.3+3437 & 14:29:18.51 & 34:37:25.8 & 1.262 & 9 & 18 & \citealp{eisenhardt2008} \\
\enddata
\tablenotetext{1}{Number of H$\alpha$ emitting members used in this analysis, which requires ``good'' stellar masses (see \S 4.3).}

\end{deluxetable}

\begin{deluxetable}{llllccclllllc}
\rotate
\tabletypesize{\scriptsize}
\tablecaption{WFC3 Grism Spectroscopic Redshifts}
\tablewidth{0pt}
\tablehead{
 \colhead{Cluster ID} & \colhead{Name} & \colhead{RA} & \colhead{Dec.} &  \colhead{$z$} & \colhead{Quality} & \colhead{Flux\footnotemark[1]} & \colhead{Fluxerr\footnotemark[1]} & \colhead{EW\footnotemark[2]} & \colhead{EWerr\footnotemark[2]} & \colhead{log M$_{*}$} & \colhead{AGN}\\
\colhead{---} & \colhead{---} & \colhead{(J2000)} & \colhead{(J2000)} &  \colhead{---} & \colhead{---} & \colhead{10$^{-17}$ ergs s$^{-1}$ cm$^{-2}$} & \colhead{10$^{-17}$ ergs s$^{-1}$ cm$^{-2}$} & \colhead{\AA} & \colhead{\AA} & \colhead{M$_{\odot}$} & \colhead{---}
}
\startdata
ISCS J1426.5+3339 & J142626.1+333827 & 14:26:26.09 & 33:38:27.6 & 1.170 & B & 79.1 & 5.3 & 843 & 81 & 10.8 & ---\\
ISCS J1426.5+3339 & J142633.8+333844 & 14:26:33.84 & 33:38:44.4 & 1.138 & B & 8.7 & 2.0 & 260 & 71 & 9.9 & ---\\
ISCS J1426.5+3339 & J142633.9+333915 & 14:26:33.87 & 33:39:15.8 & 1.173 & B & 6.5 & 1.9 & 133 & 45 & 9.5 & ---\\
ISCS J1426.5+3339 & J142632.7+333922 & 14:26:32.68 & 33:39:22.0 & 1.165 & B & 6.6 & 1.7 & 116 & 32 & 10.1 & ---\\
ISCS J1426.5+3339 & J142631.8+333914 & 14:26:31.75 & 33:39:14.8 & 1.167 & B & 9.6 & 3.0 & 109 & 34 & 10.8 & ---\\
ISCS J1426.5+3339 & J142629.0+333941 & 14:26:29.04 & 33:39:41.0 & 1.165 & B & 12.3 & 2.6 & 111 & 23 & 9.6 & ---\\
ISCS J1426.5+3339 & J142626.1+333826 & 14:26:26.12 & 33:38:26.8 & 1.170 & B & 83.7 & 4.8 & 439 & 34 & 10.8 & ---\\
ISCS J1426.5+3339 & J142629.5+333912 & 14:26:29.52 & 33:39:12.0 & 1.158 & C & 15.3 & 2.6 & 264 & 53 & 10.0 & ---\\
\enddata
\tablenotetext{1}{H$\alpha$+[N\textsc{ii}]}
\tablenotetext{2}{Observed for H$\alpha$+[N\textsc{ii}]}
\end{deluxetable}

\end{document}